# A remote response of ATP hydrolysis

Xin Liu

State Key Laboratory of Nonlinear Mechanics.

Institute of Mechanics, Chinese Academy of Sciences. Beijing 100190, China

Corresponding author:

Xin Liu

Phone: 086-010-82543970    Email: liuxin@lnm.imech.ac.cn

Fax: 086-010-82543977    Address: Beisihuanxi Road #15, Beijing China, 100190

Office: State Key Laboratory of Nonlinear Mechanics, Institute of Mechanics, Chinese Academy of Sciences.

Running title: A recycle effect of ATP hydrolyzation.

**Abstract**

ATP-hydrolysis is the basic energy source of many physiological processes, but there is a lack of knowledge regarding its biological role other than energy transfer and thermogenesis. Not all the energy released by ATP-hydrolysis could be used in powering target biological processes and functions. Partial energy dissipates into water. By validating the impact of this dissipated energy, we found that energy released by ATP hydrolysis could induce notable regulation of biomolecule's properties 100 nanometers away. Namely ATP hydrolyzation is recycled in remote biochemical property modulation.

Keywords: ATPase , recycling effect, fluorescence resonance energy transfer , large unilamellar vesicle

**Introduction**

Adenosine triphosphate (ATP) is a nucleotide that is considered to be the "molecular currency" of intracellular energy transfer. It consists in cytoplasm and nucleoplasm of every cell, and is the energy source of nearly all physiological mechanisms. The energy released and the modes of ATP hydrolyzation are different due to the participation of various enzymes. In standard state conditions, the energy stored in ATP is released by hydrolyzing an ATP molecule into adenosine diphosphate (ADP) and a phosphate ion (Pi, about 30.6 kJ/mol energy is liberated). At present, most of the existing studies focus on the utilitarian side of ATP hydrolyzation, in which energy stored in ATP is transmitted to the organism and powers molecule transportation, biochemical reaction, and etc. Only a few reports regarded other processes in the shade of this splendid energy transfer, such as the shivering thermogenesis in actomyosin and heat generating pathway in heater cells (Dejours, 1987). The energy released in ATP hydrolysis is only partially used in target biological processes, e.g. the energy conversion rate of ATPase ranges from 50% to 100% (Noji et al, 1997; Yasuda et al, 1998;Montemagno, & Bachand, 1999;Soong et al, 2000; Soong et al, 2001; Ren ,& Zhao, 2006). The "wasted" energy dissipates into the surrounding water and turns into heat eventually. It is still unclear whether such heating effect proceeds via simple energy dissipation or results in subsequent processes significant to life.

Here, we investigated this overlooked process by observing its influence on protein-protein interaction. As shown in Figure 1, subsequential functions of the "utilized" energy and the "wasted" energy were isolated from each other in a physical way; effect of the "wasted" energy was probed by its influence on protein binding ability. The $Na^+K^+$-ATPase (Skou, 1957) was planted onto LUVs (Milhiet et al, 2002;Giocondi et al, 2004), which were artificial membrane spherical shells of lipid bilayer composed of 1,2-dipalmitoyl-sn-glycero-3-phosphocholine (DPPC). Potassium ions ($K^+$) were contained in the inner solution of these hermetic LUVs. In the solution outside LUVs, there were ATP and two types of non-membrane proteins, a wild type WW-domain protein (PDBID 1WR4) and a proline-rich peptide GTPPPPYTVG. These probe proteins can recognize each other with a binding constant $K_D=75.2\pm0.4uM$ (Liu, & Zhao, 2010). As the concentration of probe complex was very small (in a scale of uM in our experiment, nearly $100^3$ nanometer$^3$ per complex), only the long-range influence of "wasted" energy could be effective in regulating the probes' biochemical properties. As such, effects of the "utilized" energy and the "wasted" energy were separated not only in topology, but also in space and distance. Fluorescein isothiocyanate (FITC) and CY3 dye had been linked to the two proteins respectively. As the two proteins combine with each other, the distance between FITC and CY3 was short enough to initiate the fluorescence resonance energy transfer (FRET) mechanism (Lakowicz, 1999). If there was any change in the recognition ability of the two proteins, we could observe it from the brightness of fluoresceins through FRET.

**Results**

$Na^+K^+$-ATPase is an enzyme that sustains cells the relatively high concentrations of potassium ions but low concentrations of sodium ions, and pumps three sodium ions out of the cell for every two potassium ions pumped in. In our experiment, impact of this well known function was astricted to the space inside LUVs and around their surfaces locally. Initially, there was not sodium ion ($Na^+$) in the outside solution of LUVs. According to the investigation by Skou (Skou, 1957), although there is a potassium independent small basal ATP hydrolysis due to the presence of magnesium, the enzyme activity is still very low in such sodium absent case; after injection of the solution of NaCl there, enzyme activity increases drastically, with a steep slope(see supplementary materials). The $Na^+K^+$-ATPase pumped $Na^+$ into the LUVs and $K^+$ outside, using the energy released by ATP-hydrolyzation. Besides energy that did this useful work, other energies dissipated into solution outside LUVs. The effect of such "wasted" energy was verified far away from LUVs by probe complex through FRET. As shown in Figure 2, an obvious decrease in the ratio of emission intensities was observed shortly after every injection of NaCl. This means an increase in the distance between fluoresceins FITC and CY3, that is, an inhibition of probe proteins recognition occurred. According to the biochemical features of this molecular motor, $Na^+K^+$-ATPase was no more active than the basal standard when KCl was injected. For this ATPase inactive instance, the aforementioned inhibition was not observed. In order to determine the effects of exotic ions on protein binding, NaCl or KCl was injected into ATPase-LUV-absent solutions, i.e. the solution outside LUVs. As shown in the supporting materials, protein recognitions were not inhibited evidently. Therefore, exotic ions did not affected protein association in our cases. For solutions used for injection, the PH values of NaCl and KCl were both 6.6, that is, acidity-alkalinities should have identical contributions to all the experiments. Therefore, the inhibition of protein recognition was neither the result of ATPase inactivation nor the contribution of exotic ions, but was induced by the effects of "wasted" energy from ATP hydrolysis.

**Materials and methods**

Reagents preparation

High-purity (>95%) proteins of the WW domain and the biotin-GTPPPPYTVG peptide were synthesized via solid phase peptide synthesis using Fmoc-Chemistry from SBS Genetech Co. Ltd. The FITC, anti-biotin CY3, ATP, $Na^+K^+$-ATPase, and DPPC were purchased from Sigma-Aldrich, Inc. WW-domain proteins were labeled with FITC, and the biotin-GTPPPPYTVG peptide was labeled with anti-biotin CY3 according to the manuals provided by Sigma-Aldrich, Inc. Suitable amounts of KCl were added to all aqueous solutions of labeled proteins, ATP, and $Na^+K^+$-ATPase to achieve a concentration of 0.15 M. These were used to achieve a balance of osmotic pressure in further experiments.

Large unilamellar vesicle preparation

A DPPC solution in chloroform/methanol 2/1 (v/v) was injected into a tube. The solvent was evaporated in a rotary evaporator at 70 $^o$C, and then stored overnight in vacuum. Multilamellar vesicles (MLVs) were prepared by adding KCl solution (0.15 M) to the tube and then sonication for 1 min at 70 $^o$C. Large unilamellar vesicles (LUVs) were obtained by extrusion of the MLV through a 0.1 um nuclepore track-etched membrane.

Validation of the long-range effects of the "wasted" energy

The solution of $Na^+K^+$-ATPase (compositions: $Na^+K^+$-ATPase 30uM, KCl 0.15M) was injected into the LUVs (1/5, v/v), and stored for 30 min at room temperature. After the $Na^+K^+$-ATPase was planted onto the LUVs, solutions of the two labeled proteins and ATP were injected into the LUVs. A 0.15 M solution of NaCl/KCl was then injected into the mixture every 2 min. Fluorescence excitation spectra were recorded every 20 s using a HITACHI F-4500 fluorospectrophotometer from Techcomp Ltd.

Temperature control

Reagents were prepared at room temperature, prohibiting any heat source nearby. Before FRET measurement, the fluorospectrophotometer had been warmed up for half-hour to achieve a stable temperature condition inside HITACHI F-4500. Before ions injection, the NaCl/KCl solution and the objects under test were loaded into fluorospectrophotometer for five minutes. Thus, there was not transient temperature change in our observation.

Reagents：
I. Exotic NaCl solution used for injection
   a) PH=6.6 at 25$^o$C,
   b) Inject 5ul every 120 seconds,
   c) Composition: NaCl (0.15M).

II.   Exotic KCl solution used for injection
   a) PH=6.6 at 25°C,
   b) Inject 5ul every 120 seconds,
   c) Composition: KCl (0.15M).
III.  The solution outside LUVs
   a) PH=5.1 at 25°C,
   b) Volume used per experiment: 300 ul,
   c) Compositions: WW-FITC (13uM), GTPPPPYTVG-biotin-CY3 (0.19uM), ATP (1.9mM), KCl (0.15M), $K_2CO_3$ (0.17mM), $Mg^+$ from ATP magnesium salt (less than 1.9mM).
IV.   The ATPase-LUV-absent solution
   a) PH=5.1 at 25°C,
   b) Volume used per experiment: 300 ul,
   c) Compositions: WW-FITC (13uM), GTPPPPYTVG-biotin-CY3 (0.19uM), ATP (1.9mM), KCl (0.15M), $K_2CO_3$ (0.17mM), $Mg^+$ from ATP magnesium salt (less than 1.9mM).
V.    The solution inside LUVs
   a) PH=6.6 at 25°C,
   b) Composition: KCl (0.15M).
VI.   Membrane and $Na^+K^+$-ATPase

   Density of $Na^+K^+$-ATPase on membrane is about 7 molecular motors per $um^2$. In each experiment, the $Na^+K^+$-ATPases are about $7\times10^{13}$ counts, can catalyze ATP with a rate about $3.3\times10^{-10}$ mol/sec, pump $10^{-9}$ mol/sec $Na^+$ into LUVs, and pump $6.7\times10^{-10}$ mol/sec $K^+$ out.

**Discussion**

It has been a consensus that ATP hydrolysis is coupled with biomolecular dynamics (Shih et al, 2000;Mizukura,& Maruta,2002). Conformational changes of molecular motor have been observed by several groups, and there are several states during the whole process of ATP hydrolysis (Abrahams et al, 1994). Whereas, span of the influence of conformational change is restricted by molecule size, thus is limited to the region around molecular motor and LUV (the typical cutoff of Lennard-Jones interaction is about 2nm). Even though the hydration shells (Frauenfelder et al, 2009) of these biomolecules are considered, the impact of biomolecular dynamics can extend only several layers of water molecule(less than 1nm). In contrast, we had observed the inhibition of protein recognition induced by ATP hydrolysis about a hundred nanometers away from $Na^+K^+$-ATPase motor. Such inhibition was distinct from the consequence of local dynamics of ATP hydrolysis. Moreover, the subsequential functions of the "utilized" energy had been astricted to the hermetic space isolated from probe protein complex. Therefore, the observed phenomenon was neither the result of local motor dynamics, nor the side effects of ion pumping, but was induced by "wasted" energy in a long range.

In water, the long range transmission of the subsequential function of "wasted" energy seems incredible. Reason of this observed "weird" effect may consist in water structure. In liquid water, a hydrogen bond forms when a hydrogen atom of a water molecule is attracted towards the oxygen atom of a neighboring water molecule. Strength of such hydrogen bond is typically 20–30 kJ/mol; there are also large numbers of weak bonds that are more infirm and easier to break. It appears that the "wasted" energy of one ATP hydrolysis can break at most one typical hydrogen bond. Whereas, there is a cooperative effect in water clusters, that is, liquid water features highly directional hydrogen-bond network structures that can associate dynamically (Luck,1998; Achim et al, 2003). If one hydrogen bond is broken, it will be easier to break a second one in the network, then a third, a fourth, and so on. Consequently, the "wasted" energy of one ATP hydrolysis could arouse large-scale water fluctuations around the position of ATP hydrolysis. As the original equilibrium of hydrogen-bond network was partly demolished, specifically, as the attraction forces suddenly decreased at the hydrolysis position, other parts of the network would shrink in size, and facilitate the avulsion of more hydrogen bonds at hydrolysis position. This could result in a global response and large-scale increase of water molecule density except of hydrolysis position. Since water structure around hydrolysis position had been broadly destroyed, the global response should last about 1 nanosecond to recover, a typical duration of system relaxation. This is much longer than the time scale $10^{-12}$ sec of the formation of hydrogen bond (Erkoc et al, 2001; Yoshioki, 2004).

The typical binding free energy of protein-protein interaction ranges from -25 to -80 kJ/mol ( Keskin et al, 2008), and that of the present probe complex is about -24 kJ/mol. It appears that the "wasted" energy (about 10 kJ/mol) is not large enough to induce a direct inhibition of protein-protein recognition. However, the protein-protein interaction is not accomplished at one stroke, but achieved step by step, with many metastable states in its energy landscape. Therefore, "wasted" energy could play its part in aspect of free energy. On the other hand, recent investigations have shown that biological molecules are sensitive to changes in molecular mechanical properties. As the basis of side-chain fluctuations, loops movement at active site, and structural exchanges and rearrangements, molecular mechanical properties have been proposed to be vital for molecule's function, internal motion, and evolution (Smock, & Gierasch, 2009; Tokuriki, & Tawfik, 2009; Engler et al, 2009). Some homologous proteins can even be designed entirely based on dynamism (Liu, & Zhao, 2010). Therefore, a dynamism impacting event can result in profound influences on molecule's biochemical properties. The aforementioned global increase of water density could boost up the squeezing effect (Liu, & Zhao, 2010) on remote biomolecules, increase the amount of hydrogen bond donor in water, drive the dynamic equilibrium of hydrogen-bond network to be water preference, and remodel hydration shells (Frauenfelder et al, 2009) around these biomolecules. This could inevitably affect molecular dynamism, result in notable regulations to their biochemical properties, such as initial biomolecule interactions, backbone dynamic fluctuations, and so on.

The ATP-based remote regulation completes the picture of ATP hydrolyzation. It could be a remarkable biological function of ATP-hydrolyzation. As observed in our experiment, such a process could destroy weak interfaces of biomolecular complex. Only "good" interaction modes were maintained under such vigorous filtration. Similar mechanisms could also reduce the "unfavorable" folding process, which is inefficient in producing correctly folding nuclei, such that the native-folded state of a protein can dominate the outcome of various folding pathway (Dobson, 2004). This might be one of the long-range housekeeping mechanisms that determine the correct physiological process, and could be deemed a general mechanism in integrating numerous biomolecules into a live organism. Due to such regulation, biomolecules are no longer individual units, but parts of a united system that can interact, cooperate, and even resonate.

As a basic regulation mechanism, such remote response can be potentially involved in the processes of many nanoscale cooperations of life. For example, muscles are well known for their cooperative feature, where remote response of ATP hydrolysis may participate in the formation of different contraction mechanisms between skeletal muscle and smooth muscle. Tropomyosin is a protein about 40nm in length. It associates with actin in muscle fibers, overlays myosin binding sites of actin, and prevents mysion-actin binding in resting muscle. In muscle contraction, the tropomyosin changes its conformation to expose the binding groove of actin, such that myosin heads can access the binding sites of actin, and induce ATP hydrolysis and filament sliding. As thus, the conformation change of tropomyosin is vital to muscle contraction. Many tropomyosins along the actin change their conformation cooperatively to induce a collaborative myosin binding and ATP hydrolysis. But, for skeletal muscle and smooth muscle, the modes of conformation change cooperation are different. In skeletal muscle, the action potential induces the release of calcium. Then the calcium binds to troponin-tropomyosin-actin complex, resulting in conformation change of tropomyosin. In every contraction, the tropomyosins along actin unlock cooperatively. Firstly, this is due to the simultaneous calcium release responding to action potential. Secondly, one ATP hydrolysis could decrease the tropomyosin-actin binding in a scale of nanometers, just as that of our experiment, and make the binding sites expose more synchronous. While in smooth muscle, the status is different. On one hand, there is less action potential than that of skeletal muscle. On the other hand, it is intriguing that the tropomyosin filament moves closer to the outer actin domain in the $Ca^{2+}$-calmodulin activating condition, by an azimuthal rotation of 20~25°, corresponding to a shrinkage of 15~20Å. This effect is not observed on striated muscle activation (Perry, 2001). It induces an increase of the effect of remote ATP hydrolysis response, and enhances the significance of the second factor in unlocking cooperation. Namely, it seems that there is a shift of the importance of the two factors.

**Figure legends**

Figure 1. Experiment sketch in validating the effect of "wasted" energy of ATP hydrolyzation. After injection of NaCl solution (Step I), the $Na^+K^+$-ATPase hydrolyzed ATP molecules into ADP and Pi, and pumped the $Na^+/K^+$ in/out of the hermetic LUV solution (Step II). Besides accomplishing this useful work, effect of "wasted" energy was verified far away from the hydrolysis position (Step III). The influence of such effects in protein recognition was observed by the change of fluorescein brightness through FRET.

Figure 2. FRET results induced by long-range effects of the "wasted" energy of ATP hydrolyzation. When the distance of two fluorescence probes become longer, there is a decrease in the amount of red light emitted by CY3 and an increase of green light emitted by FITC. We characterized the distance by the ratio of the emission intensities (REI) between CY3 and FITC. Emission intensities at wavelengths of 520 and 570nm were analyzed; these corresponded to FITC and CY3, respectively. A decrease in the ratio means an increase of distance. Solution injections were performed at 0, 120, and 240 s. The fluorescence excitation spectra were measured every 20 sec. Upper: The results of injection of NaCl solution. Shortly after the injections, we observed decreases of REI much larger than the levels of squared deviations from the mean of background REIs (shown by red dotted lines, estimated from four points before the next injection). Such a decrease in red light emission was also observed in the insertion, where two examples of fluorescence excitation spectra were reported. Lower: The results of injection of KCl solution. In these cases, $Na^+K^+$-ATPase was no more active than the basal standard; there was not decrease in red light emission compared to the upper cases. The REIs of the no-injection cases is reported in the insertion. The results show that REIs decreased linearly with time. Therefore, the systems were stable while experiments were carried out. All experiments were performed twice independently, and similar results were obtained.

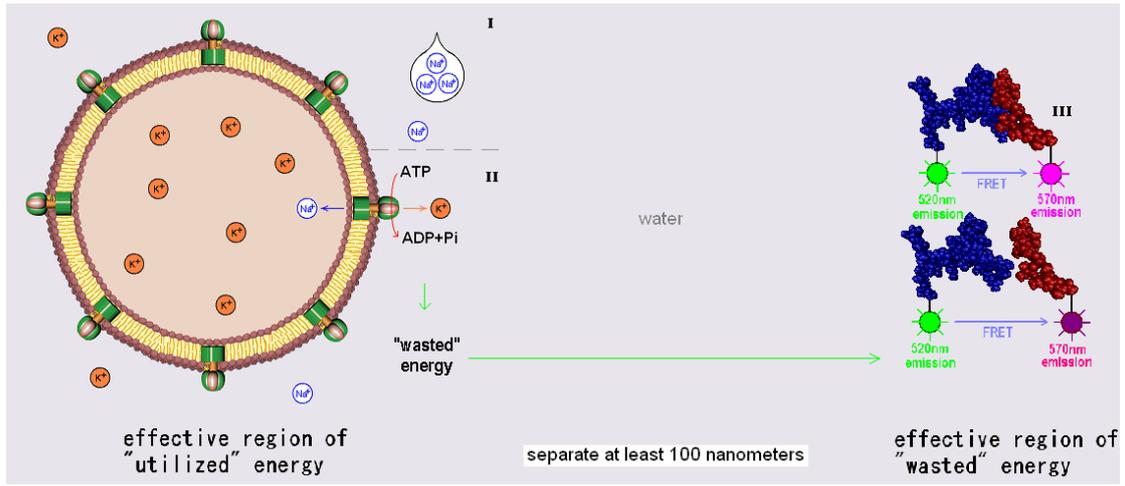

Figure 1

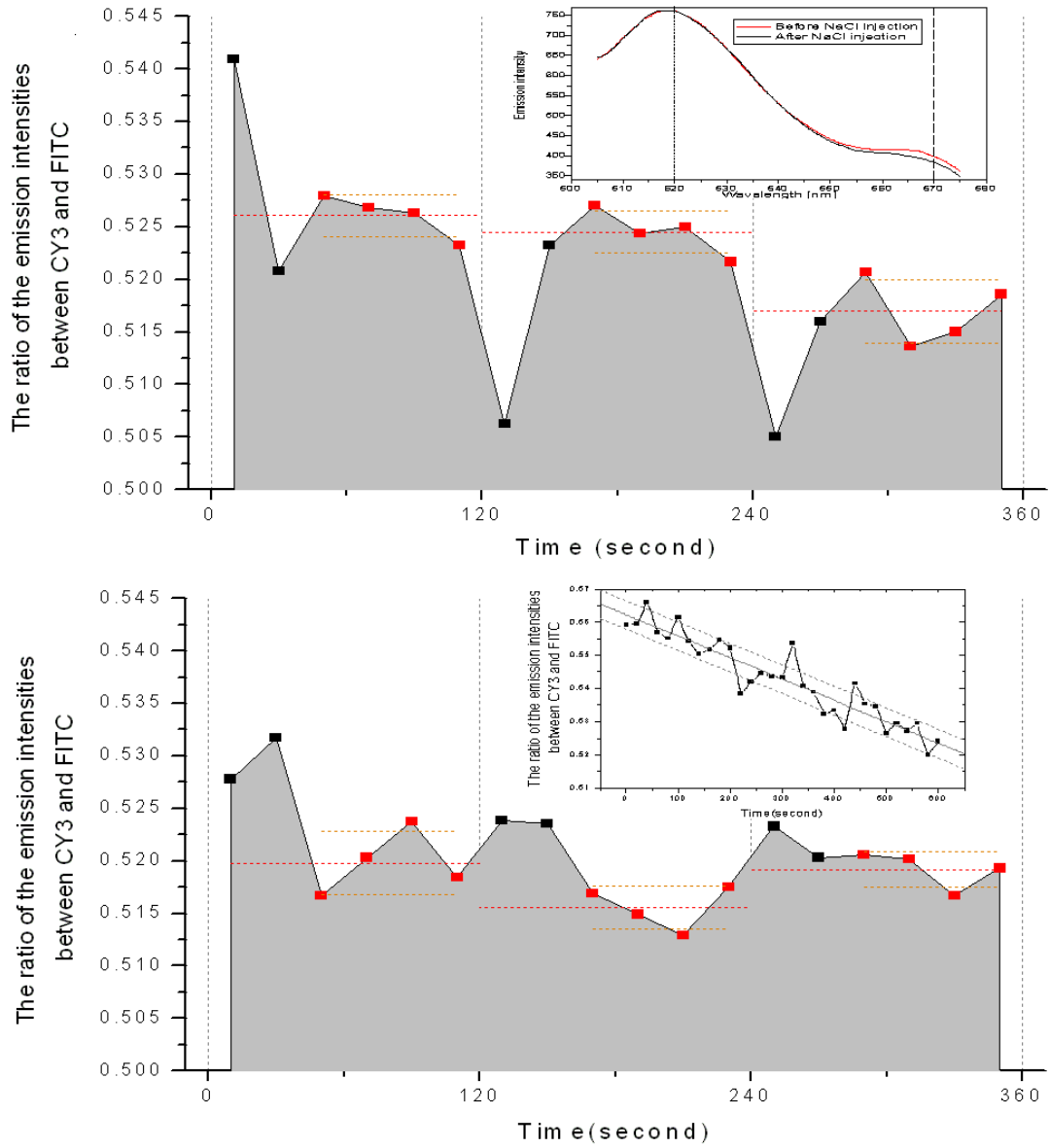

Figure 2